\documentclass[12pt]{article}
\usepackage{amsmath}
\usepackage{graphicx}
\usepackage{amsfonts}
\usepackage{amssymb}
\begin{document}
\noindent{\small CITUSC/01-024\hfill\hfill hep-th/0106157 }\linebreak
{\small NSF-ITP-01-70\hfill}

{\vskip1.5cm}

\begin{center}
{\Large \textbf{Map of Witten's }}$\star$ {\Large \textbf{to Moyal's }}$\star$

\bigskip

{\vskip0.5cm}

\textbf{Itzhak Bars}{\footnote{This research was partially supported by the US
Department of Energy under grant number DE-FG03-84ER40168 and by the National
Science Foundation under grant Nos. PHY99-07949 and INT97-24831.}}

{\vskip0.5cm}

\textsl{CIT-USC Center for Theoretical Physics \& Department of Physics}

\textsl{University of Southern California,\ Los Angeles, CA 90089-2535, USA}

{\vskip0.5cm}

\textbf{Abstract}
\end{center}

It is shown that Witten's star product in string field theory, defined as the
overlap of half strings, is equivalent to the Moyal star product involving the
relativistic phase space of even string modes. The string field $\psi
_{A}\left(  x^{\mu}\left[  \sigma\right]  \right)  $ can be rewritten as a
phase space field of the even modes $A\left(  x_{2n}^{\mu},x_{0},p_{2n}^{\mu
}\right)  ,$ where $x_{2n}^{\mu}$ are the positions of the even string modes,
and $p_{2n}^{\mu}$ are related to the Fourier space of the odd modes
$x_{2n+1}^{\mu}$ up to a linear transformation. The $p_{2n}^{\mu}$ play the
role of conjugate momenta for the even modes $x_{2n}^{\mu}$ under the string
star product. The split string formalism is used in the intermediate steps to
establish the map from Witten's $\star$-product to Moyal's $\star$-star
product$.$ An ambiguity related to the midpoint in the split string formalism
is clarified by considering odd or even modding for the split string modes,
and its effect in the Moyal star product formalism is discussed. The
noncommutative geometry defined in this way is technically similar to the one
that occurs in noncommutative field theory, but it includes the timelike
components of the string modes, and is Lorentz invariant. This map could be
useful to extend the computational methods and concepts from noncommutative
field theory to string field theory and vice-versa.

\bigskip

\section{Moyal's star product in half Fourier space}

A long time ago deformation quantization was developed as a method for
studying quantum mechanics. The methods for establishing the correspondence
between deformation quantization and the traditional formulation of quantum
mechanics was developed by Weyl, Wigner, Moyal and many others \cite{weyl} -
\cite{moyal}, leading the way to the modern ideas of noncommutative geometry
\cite{douglas}. Using Weyl's correspondence \cite{weyl} one establishes a map
between functions of operators $\hat{A}\left(  \hat{x}^{M},\hat{p}_{M}\right)
$ acting in a Hilbert space and their image in phase space $A\left(
x^{M},p_{M}\right)  .$ To preserve the quantum rules, phase space functions
must be multiplied with each other by using the associative, noncommutative,
Moyal star product
\begin{equation}
\left(  A\star B\right)  \left(  x,p\right)  =A\left(  x,p\right)
\,e^{\frac{i\hbar}{2}\left(  \frac{\overleftarrow{\partial}}{\partial x^{M}%
}\frac{\overrightarrow{\partial}}{\partial p_{M}}-\frac{\overleftarrow
{\partial}}{\partial p_{M}}\frac{\overrightarrow{\partial}}{\partial x^{M}%
}\right)  }B\left(  x,p\right)  \label{moyalstar}%
\end{equation}
(or its deformed generalizations). Note that a spacetime metric does not enter
in this expression because all positions have upper spacetime indices and all
momenta have lower spacetime indices. If we use the notation $X^{m}%
\equiv\left(  x^{M},p_{M}\right)  $ with a single index $m$ that takes
$D\times2$ values, the Moyal star product takes a form which is more familiar
in the recent physics literature on noncommutative geometry
\begin{equation}
\left(  A\star B\right)  \left(  X\right)  =A\left(  X\right)  \,e^{\frac
{i}{2}\theta^{mn}\frac{\overleftarrow{\partial}}{\partial X^{m}}%
\frac{\overrightarrow{\partial}}{\partial X^{n}}}B\left(  X\right)  ,
\end{equation}
where $\theta^{mn}=\hbar\delta_{\,\,N}^{M}\,\varepsilon_{ij},$ with $i=\left(
1,2\right)  $ referring to $\left(  x,p\right)  $ respectively, and
$\varepsilon_{ij}$ the antisymmetric Sp$\left(  2,R\right)  $ invariant
metric. Henceforth we will set $\hbar=1$ for simplicity. The star commutator
between any two phase space fields is defined by $\left[  A,B\right]  _{\star
}\equiv A\star B-B\star A.$ The phase space coordinates satisfy $\left[
X^{m},X^{n}\right]  _{\star}=i\theta^{mn},$ which is equivalent to the
Heisenberg algebra for $\left(  x^{M},p_{M}\right)  $%
\begin{equation}
\left[  x^{M},x^{N}\right]  _{\star}=\left[  p_{M},p_{N}\right]  _{\star
}=0,\quad\left[  x^{M},p_{N}\right]  _{\star}=i\delta_{N}^{M}.
\end{equation}

Let us now consider the Fourier transform in the momentum variables $p_{M}$.
We will call this ``half-Fourier space'' since only one of the noncommutative
variables is being Fourier transformed. So, the transform of $A\left(
x^{M},p_{M}\right)  $ is a bi-local function $\psi_{A}\left(  x^{M}%
,y^{M}\right)  $ in position space, but we will write it in the form
$\tilde{A}\left(  l^{M},r^{M}\right)  \equiv\psi_{A}\left(  x^{M}%
,y^{M}\right)  $ where
\begin{equation}
l^{M}=x^{M}+\frac{y^{M}}{2},\quad r^{M}=x^{M}-\frac{y^{M}}{2}. \label{lr}%
\end{equation}
Thus, we define
\begin{equation}
A\left(  x^{M},p_{M}\right)  =\int d^{D}y\,e^{-iy^{M}p_{M}}\,\tilde{A}\left(
x^{M}+\frac{y^{M}}{2},x^{M}-\frac{y^{M}}{2}\right)  =\int d^{D}y\,e^{-iy^{M}%
p_{M}}\,\psi_{A}\left(  x^{M},y^{M}\right)  . \label{ffourier}%
\end{equation}
The Moyal star product of two functions $\left(  A\star B\right)  \left(
x,p\right)  =C\left(  x,p\right)  $ may now be evaluated in terms of these
integral representations. The result can be written in terms of the Fourier
transform of $C\left(  x,p\right)  \rightarrow\tilde{C}\left(  l^{M}%
,r^{M}\right)  =\psi_{C}\left(  x^{M},y^{M}\right)  $
\begin{equation}
\left(  A\star B\right)  \left(  x,p\right)  =\int d^{D}y\,e^{-ip\cdot
y}\,\tilde{C}\left(  x+\frac{y}{2},x-\frac{y}{2}\right)  .
\end{equation}
One finds that $\tilde{C}\left(  l^{M},r^{M}\right)  $ is related to
$\tilde{A}\left(  l^{M},r^{M}\right)  $ and $\tilde{B}\left(  l^{M}%
,r^{M}\right)  $ by a matrix-like multiplication with continuous indices
\begin{equation}
\tilde{C}\left(  l^{M},r^{M}\right)  =\int d^{D}y\,\tilde{A}\left(
l^{M},y^{M}\right)  \tilde{B}\left(  y^{M},r^{M}\right)  . \label{matrix}%
\end{equation}
This is verified explicitly by the following steps
\begin{align}
A\star B  &  =\int d^{D}yd^{D}y^{\prime}\,\,\tilde{A}\left(  x+\frac{y}%
{2},x-\frac{y}{2}\right)  \,e^{-iy\cdot p}e^{\frac{i}{2}\left(  \frac
{\overleftarrow{\partial}}{\partial x}\frac{\overrightarrow{\partial}%
}{\partial p}-\frac{\overleftarrow{\partial}}{\partial p}\frac{\overrightarrow
{\partial}}{\partial x}\right)  }e^{-iy^{\prime}\cdot p}\,\tilde{B}\left(
x+\frac{y^{\prime}}{2},x-\frac{y^{\prime}}{2}\right) \nonumber\\
&  =\int d^{D}yd^{D}y^{\prime}\,e^{-iy\cdot p}\,\tilde{A}\left(  x+\frac{y}%
{2},x-\frac{y}{2}\right)  e^{\frac{1}{2}\left(  \frac{\overleftarrow{\partial
}}{\partial x}y^{\prime}-y\frac{\overrightarrow{\partial}}{\partial x}\right)
}\,\tilde{B}\left(  x+\frac{y^{\prime}}{2},x-\frac{y^{\prime}}{2}\right)
e^{-iy^{\prime}\cdot p}\nonumber\\
&  =\int d^{D}yd^{D}y^{\prime}\,e^{-i\left(  y+y^{\prime}\right)  \cdot
p}\,\tilde{A}\left(  x+\frac{y+y^{\prime}}{2},x-\frac{y-y^{\prime}}{2}\right)
\,\tilde{B}\left(  x-\frac{y-y^{\prime}}{2},x-\frac{y+y^{\prime}}{2}\right)
\nonumber\\
&  =\int d^{D}y_{+}d^{D}y_{-}\,e^{-ip\cdot y_{+}}\,\tilde{A}\left(
x+\frac{y_{+}}{2},y_{-}\right)  \,\tilde{B}\left(  y_{-},x-\frac{y_{+}}%
{2}\right) \nonumber\\
&  =\int d^{D}y_{+}\,e^{-ip\cdot y_{+}}\,\tilde{C}\left(  x+\frac{y_{+}}%
{2},x-\frac{y_{+}}{2}\right)  \, \label{proof}%
\end{align}
where, in the second line the derivatives with respect to $p$ are evaluated,
in the third line the translation operators $e^{\frac{1}{2}\frac
{\overleftarrow{\partial}}{\partial x}y^{^{\prime}}},$ $e^{-\frac{1}{2}%
y\frac{\overrightarrow{\partial}}{\partial x}}$ are applied on the $x$
coordinates on the left and right respectively, in the fourth line one defines
$y_{+}=y+y^{\prime}$ and $y_{-}=x-\left(  y-y^{\prime}\right)  /2$, and
finally the $y_{-}$ integration is performed.

Hence, in our notation, the Moyal star product in Fourier space is equivalent
to infinite matrix multiplication with the rules of Eq.(\ref{matrix}): the
right variable of $\tilde{A}$ is identified with the left variable of
$\tilde{B}$ and then integrated. Eq.(\ref{matrix}) is the key observation for
establishing a direct relation between Witten's star-product and Moyal's
star-product as we will see below.

The fact that the Moyal star-product is related to some version of matrix
multiplication is no surprise, as by now a few versions of matrix
representations have been used in the physics literature. The one used here is
straightforward: after using the Weyl correspondence to derive an operator
$\hat{A}$ from the function $A\left(  x,p\right)  ,$ the matrix representation
$A\left(  l,r\right)  $ is nothing but the matrix elements of the operator
$\hat{A}$ in position space: $A\left(  l,r\right)  =\langle l|\hat{A}|r\rangle.$

\section{Witten's star product in split string space}

We will show that the continuous matrix representation of the Moyal star
product of the previous section is in detail related to Witten's star product
in string field theory. The rough idea is to replace the points $l^{M},r^{M}$
by left and right sides of a string $x^{\mu}\left[  \sigma\right]  =l^{\mu
}\left[  \sigma\right]  \oplus r^{\mu}\left[  \sigma\right]  $ (with $l,r$
defined relative to the midpoint at $\sigma=\pi/2$). If we consider the fields
of two strings $\tilde{A}\left(  l_{1}^{\mu}\left[  \sigma\right]  ,r_{1}
^{\mu}\left[  \sigma\right]  \right)  $ and $\tilde{B}\left(  l_{2}^{\mu
}\left[  \sigma\right]  ,r_{2}^{\mu}\left[  \sigma\right]  \right)  $, then
Witten's string star product is formally given by the functional integral
\cite{witten}
\begin{equation}
\tilde{C}\left(  l_{1}^{\mu}\left[  \sigma\right]  ,r_{2}^{\mu}\left(
\sigma\right)  \right)  =\int\left[  dz\right]  \,\tilde{A}\left(  l_{1}^{\mu
}\left[  \sigma\right]  ,z^{\mu}\left[  \sigma\right]  \right)  \,\tilde
{B}\left(  z^{\mu}\left[  \sigma\right]  ,r_{2}^{\mu}\left[  \sigma\right]
\right)  , \label{wittenstar}%
\end{equation}
where $z^{\mu}\left[  \sigma\right]  =r_{1}^{\mu}\left[  \sigma\right]
=l_{2}^{\mu}\left[  \sigma\right]  $ corresponds to the overlap of half of the
first string with half of the second string, and $\tilde{C}\left(  l_{1}^{\mu
}\left[  \sigma\right]  ,r_{2}^{\mu}\left(  \sigma\right)  \right)  $ is the
field describing the joined half strings as a new full string $x_{3}^{\mu
}\left[  \sigma\right]  =l_{1}^{\mu}\left[  \sigma\right]  \oplus r_{2}^{\mu
}\left(  \sigma\right)  $. Considering the close analogy to Eq.(\ref{matrix}),
morally we anticipate to be able to rewrite Witten's star product as a Moyal
star-product of the form (\ref{moyalstar}) in a larger space. In the remainder
of this paper the details of this map will be clarified, and will be shown
that Witten's $\star$ is indeed Moyal's $\star$ in \textit{half of a
relativistic phase space} of the full string, involving only the even modes
$(x_{2n},p_{2n})$ or only the odd modes $(x_{2n-1},p_{2n-1})$.

Our result may be summarized as follows: Define the Fourier transform of the
string field $\psi_{A}(x_{0},x_{2n},x_{2n-1})\equiv\tilde{A}\left(  l_{1}%
^{\mu}\left[  \sigma\right]  ,r_{1}^{\mu}\left[  \sigma\right]  \right)  $ in
the odd modes only as follows%

\begin{equation}
A(x_{x_{2n}},x_{0},p_{2n})=\int(\prod_{n=1}^{\infty}dx_{2n-1})\ \psi_{A}%
(x_{0},x_{2n},x_{2n-1})\ exp\left(  {-2i\sum_{k,l=1}^{\infty}p_{2k}^{\mu
}(\mathcal{T}_{2k,2l-1})x_{2l-1}^{\nu}\eta_{\mu\nu}}\right)  \label{fourier}%
\end{equation}
where $\mathcal{T}_{2k,2l-1}$ is a matrix to be defined below. Then Witten's
star product (\ref{wittenstar}) for two string fields $(\psi_{A}\star
_{witten}\psi_{B})(x_{0},x_{2n},x_{2n-1})$ is equivalent to the Moyal star
product for their Fourier transformed fields $(A\star B)(x_{2n},x_{0},p_{2n})$
with the usual definition of the Moyal star product involving the phase space
of \textit{only the even modes}%

\begin{equation}
\star\equiv\exp\left(  \frac{i}{2}\sum_{n=1}^{\infty}\eta^{\mu\nu}\left(
\frac{\overleftarrow{\partial}}{\partial x_{2n}^{\mu}}\frac{\overrightarrow
{\partial}}{\partial p_{2n}^{\nu}}-\frac{\overleftarrow{\partial}}{\partial
p_{2n}^{\nu}}\frac{\overrightarrow{\partial}}{\partial x_{2n}^{\mu}}\right)
\right)  . \label{newstar}%
\end{equation}
Furthermore, the definition of trace is the phase space integration%

\begin{equation}
Tr(\psi)=TrA\equiv\int\prod_{n=1}^{\infty}\frac{dx_{2n}dp_{2n}}{2\pi
}\ A(x_{2n},x_{0},p_{2n}). \label{trace}%
\end{equation}
In the definitions of both the star $(A\star B)(x_{2n},x_{0},p_{2n})$ and
trace $TrA$ either the center of mass mode $x_{0}$ or the midpoint mode
$\bar{x}\equiv x(\frac{\pi}{2})=x_{0}+\sqrt{2}\sum_{n=1}^{\infty}%
x_{2n}(-1)^{n}$ is held fixed while taking derivatives or doing integration
with respect to the $x_{2n}$ modes. This is precisely related to the midpoint
ambiguity in the split string formalism which will be clarified in the next section.

Witten's star product (\ref{wittenstar}) is more carefully defined in the
split string formalism which was developed sometime ago \cite{bordes}%
\cite{abdur} and was used in recent studies of string field theory
\cite{rastsenzwi} \cite{grosstaylor}\cite{kawano}. As mentioned in the
previous paragraph there is a dilemma involving the midpoint which so far has
remained obscure in the literature. We will address this issue in the next
section by considering the options that are available in the formulation of
the split string formalism, namely odd versus even modding of the split string
modes \cite{bordes}\cite{abdur}. The choice affects the definition of the star
product within the split string formalism. In turn this choice is related to
whether the center of mass mode $x_{0}$ or the midpoint mode $\bar{x}$ is held
fixed as described in the previous paragraph. In this section we begin with
the odd modding that has been used in the recent literature.

The open string position modes are identified as usual by the expansion
$x(\sigma)=x_{0}+\sqrt{2}\sum_{n=1}^{\infty}x_{n}{\cos}\left(  n\sigma\right)
{\ }$(with $0\leq\sigma\leq\pi$), where $x_{0}$ is the center of mass position
of the string. We will omit the spacetime index $\mu$ on all vectors whenever
there is no confusion. The position of the midpoint is given by $\bar{x}\equiv
x(\frac{\pi}{2})=x_{0}+\sqrt{2}\sum_{n=1}^{\infty}x_{n}\cos\left(  \frac{n\pi
}{2}\right)  .$ Therefore, instead of $x_{0}$ we may use $\bar{x}$ as the
independent degree of freedom and write the following mode expansions for the
full string $x(\sigma),$ as well as for the left side $l\left(  \sigma\right)
\equiv\left\{  x(\sigma)\,|\,0\leq\sigma\leq\frac{\pi}{2}\right\}  $ and the
right side $r\left(  \sigma\right)  \equiv\left\{  x(\pi-\sigma)\,|\,0\leq
\sigma\leq\frac{\pi}{2}\right\}  $ of the same string
\begin{align}
x(\sigma)  &  =x_{0}+\sqrt{2}\sum_{n=1}^{\infty}x_{n}{\cos}\left(
n\sigma\right)  ,\quad0\leq\sigma\leq\pi,\\
&  =\bar{x}+\sqrt{2}\sum_{n=1}^{\infty}x_{n}\left(  {\cos}\left(
n\sigma\right)  -\cos\left(  \frac{n\pi}{2}\right)  \right) \\
l\left(  \sigma\right)   &  =\bar{x}+\sqrt{2}\sum_{n=1}^{\infty}l_{2n-1}%
\cos\left(  \left(  2n-1\right)  \sigma\right)  ,\quad0\leq\sigma\leq\frac
{\pi}{2},\\
r\left(  \sigma\right)   &  =\bar{x}+\sqrt{2}\sum_{n=1}^{\infty}r_{2n-1}%
\cos\left(  \left(  2n-1\right)  \sigma\right)  ,\quad0\leq\sigma\leq\frac
{\pi}{2}.
\end{align}
These mode expansions are obtained by imposing Neumann boundary conditions at
the ends of the string $\partial_{\sigma}x|_{0,\pi}=\partial_{\sigma}%
l|_{0}=\partial_{\sigma}r|_{0}=0,$ and Dirichlet boundary conditions at the
midpoint $x(\frac{\pi}{2})=l(\frac{\pi}{2})=r(\frac{\pi}{2})=\bar{x}.$ Using
the completeness and orthogonality of the trigonometric functions in these
expansions one can easily extract the relationship between the left/right
modes and the full string modes
\begin{align*}
l_{2n-1}  &  =\frac{2\sqrt{2}}{\pi}\int_{0}^{\frac{\pi}{2}}d\sigma\,\left(
l\left(  \sigma\right)  -\bar{x}\right)  \cos\left(  2n-1\right)  \sigma
=\frac{2\sqrt{2}}{\pi}\int_{0}^{\frac{\pi}{2}}d\sigma\,\left(  x\left(
\sigma\right)  -\bar{x}\right)  \cos\left(  2n-1\right)  \sigma\\
r_{2n-1}  &  =\frac{2\sqrt{2}}{\pi}\int_{0}^{\frac{\pi}{2}}d\sigma\,\left(
r\left(  \sigma\right)  -\bar{x}\right)  \cos\left(  2n-1\right)  \sigma
=\frac{2\sqrt{2}}{\pi}\int_{0}^{\frac{\pi}{2}}d\sigma\,\left(  x\left(
\pi-\sigma\right)  -\bar{x}\right)  \cos\left(  2n-1\right)  \sigma\\
x_{n\neq0}  &  =\frac{\sqrt{2}}{\pi}\int_{0}^{\pi}d\sigma\,x(\sigma
)\,\cos\left(  n\sigma\right)  =\frac{\sqrt{2}}{\pi}\int_{0}^{\frac{\pi}{2}%
}d\sigma\left[  l\left(  \sigma\right)  +\left(  -1\right)  ^{n}r\left(
\sigma\right)  \right]  \cos\left(  n\sigma\right)  .
\end{align*}
The result is
\begin{align}
x_{2n-1}  &  =\frac{1}{2}\left(  l_{2n-1}-r_{2n-1}\right) \\
x_{2n\neq0}  &  =\frac{1}{2}\sum_{m=1}^{\infty}\mathcal{T}_{2n,2m-1}\left(
l_{2m-1}+r_{2m-1}\right) \\
x_{0}  &  =\bar{x}+\frac{1}{4}\sum_{m=1}^{\infty}\mathcal{T}_{0,2m-1}\left(
l_{2m-1}+r_{2m-1}\right)
\end{align}
where
\begin{align}
\mathcal{T}_{2n,2m-1}  &  =\frac{4}{\pi}\int_{0}^{\frac{\pi}{2}}d\sigma
\cos\left(  \left(  2n\right)  \sigma\right)  \cos\left(  \left(  2m-1\right)
\sigma\right) \\
&  =\frac{2\left(  -1\right)  ^{m+n+1}}{\pi}\left(  \frac{1}{2m-1+2n}+\frac
{1}{2m-1-2n}\right)  . \label{t0}%
\end{align}
The inverse relations are
\begin{align}
l_{2m-1}  &  =x_{2m-1}+\sum_{n=1}^{\infty}\mathcal{R}_{2m-1,2n}x_{2n}%
\label{l}\\
\bar{x}  &  =x_{0}+\sqrt{2}\sum_{n=1}^{\infty}\left(  -1\right)  ^{n}x_{2n}\\
r_{2m-1}  &  =-x_{2m-1}+\sum_{n=1}^{\infty}\mathcal{R}_{2m-1,2n}x_{2n}
\label{r}%
\end{align}
where
\begin{align}
\mathcal{R}_{2m-1,2n}  &  =\frac{4}{\pi}\int_{0}^{\frac{\pi}{2}}d\sigma
\cos\left(  2m-1\right)  \sigma\,\left[  \cos2n\sigma-\left(  -1\right)
^{n}\right] \\
&  =\mathcal{T}_{2n,2m-1}-\left(  -1\right)  ^{n}\mathcal{T}_{0,2m-1}\\
&  =\frac{4n\left(  -1\right)  ^{n+m}}{\pi\left(  2m-1\right)  }\left(
\frac{1}{2m-1+2n}-\frac{1}{2m-1-2n}\right)  .
\end{align}
We note that $\mathcal{T}_{0,2m-1}$ is given by Eq.(\ref{t0}), but it also
satisfies $\mathcal{T}_{0,2n-1}=-2\sum_{k=1}^{\infty}\left(  -1\right)
^{k}\mathcal{T}_{2k,2n-1}.$ It must be mentioned that $\mathcal{R}_{2k-1,2m}$
is the inverse of $\mathcal{T}_{2m,2n-1}$ on both sides
\begin{equation}
(\mathcal{R}\mathcal{T})_{2n-1,2k-1}=\delta_{n,k},\qquad(\mathcal{T}%
\mathcal{R})_{2m,2l}=\delta_{m,l}. \label{inverse}%
\end{equation}

In the split string notation the string field is $\psi_{A}(x_{0}%
,x_{2n},x_{2n-1})\equiv\tilde{A}\left(  \left\{  l_{2n-1}\right\}  ,\bar
{x},\left\{  r_{2n-1}\right\}  \right)  .$ Note that the midpoint $\bar{x}$ is
treated as the independent degree of freedom rather than the center of mass
mode $x_{0}$. Witten's star product takes the form (no integration over
$\bar{x}$)
\begin{equation}
\tilde{C}\left(  \left\{  l_{2n-1}\right\}  ,\bar{x},\left\{  r_{2n-1}%
\right\}  \right)  =\int\,\tilde{A}\left(  \left\{  l_{2n-1}\right\}  ,\bar
{x},\left\{  z_{2n-1}\right\}  \right)  \,\tilde{B}\left(  \left\{
z_{2n-1}\right\}  ,\bar{x},\left\{  r_{2n-1}\right\}  \right)  \prod
_{k}dz_{2k-1}. \label{wit}%
\end{equation}
By analogy to section-1 we see that we should compare $\left(  \left\{
l_{2m-1}^{\mu}\right\}  ,\left\{  r_{2m-1}^{\mu}\right\}  \right)  $ to
$\left(  l^{M},r^{M}\right)  $ and therefore via Eqs.(\ref{lr},\ref{l}%
,\ref{r}) we should establish the following correspondence
\begin{equation}
\sum_{n=1}^{\infty}\mathcal{R}_{2m-1,2n}x_{2n}^{\mu}\sim x^{M};\quad
x_{2m-1}^{\mu}\sim\frac{y^{M}}{2}.
\end{equation}
This suggests that we define a Fourier transform in twice the odd modes
$\left(  2x_{2m-1}^{\mu}\right)  \sim y^{M}$ to obtain the string field in
phase space. The Fourier parameters would play the role of conjugate momenta
to the following combination of even modes $\mathcal{R}x_{e}\equiv\sum
_{n=1}^{\infty}\mathcal{R}_{2m-1,2n}x_{2n}^{\mu}\sim x^{M}.$ We will use the
symbol $\mathcal{R}x_{e}$ as a short hand notation to indicate that the even
modes (denoted by the subscript $e$) are transformed by the matrix
$\mathcal{R}$. Therefore it is convenient to choose the Fourier parameters in
the combination $p_{e}\mathcal{T}\equiv\left\{  \sum_{n=1}^{\infty}p_{2n}%
^{\mu}\mathcal{T}_{2n,2m-1};\quad m=1,2,\cdots\right\}  $ $\sim p_{M}.$ We
will also use $x_{odd}$ as a short hand notation for $x_{odd}=\left\{
x_{2n-1}^{\mu}\right\}  $. Then we define the string field in phase space by
complete analogy to Eq.(\ref{ffourier})
\begin{equation}
A\left(  \mathcal{R}x_{e},\bar{x},p_{e}\mathcal{T}\right)  =\int\left[
dx_{odd}\right]  \,\exp\left(  -i\left(  p_{e}\mathcal{T}\cdot2x_{odd}\right)
\right)  \,\tilde{A}\left(  \mathcal{R}x_{e}+x_{odd},\bar{x},\mathcal{R}%
x_{e}-x_{odd}\right)  \label{stringFourier}%
\end{equation}
The right hand side is the same expression given in (\ref{fourier}) since
$\tilde{A}\left(  \mathcal{R}x_{e}+x_{odd},\bar{x},\mathcal{R}x_{e}%
-x_{odd}\right)  =\psi_{A}\left(  x_{0},x_{2n},x_{2n-1}\right)  $. This
construction guarantees that Witten's star product in the split string
notation of Eq.(\ref{wit}) will be precisely reproduced by a Moyal star
product in which $\mathcal{R}x_{e}$ and $p_{e}\mathcal{T}$ are the conjugate
phase space variables. This is seen by retracing the steps of the computations
that lead to Eq.(\ref{proof}). Therefore, the Moyal star should satisfy
$\left[  \mathcal{R}x_{e},p_{e}\mathcal{T}\right]  _{\star}\sim i,$ or in more
detail
\begin{equation}
\left[  \left(  \sum_{n=1}^{\infty}\mathcal{R}_{2k-1,2n}x_{2n}^{\mu}\right)
,\left(  \sum_{m=1}^{\infty}\mathcal{T}_{2m,2l-1}p_{2m}^{\nu}\right)  \right]
_{\star}=i\delta_{k,l}\eta^{\mu\nu}.
\end{equation}
However, using the fact that $\mathcal{R}$ and $\mathcal{T}$ are each other's
inverses we see that this is equivalent to the simple Heisenberg commutation
relations under the Moyal star product given in Eq.(\ref{newstar})%

\begin{equation}
\lbrack x_{2m}^{\mu},p_{2n}^{\nu}]_{\star}=i\eta^{\mu\nu}. \label{heisenberg}%
\end{equation}
Therefore, the Witten star-product reduces just to the usual Moyal product in
the phase space of only the even modes. Since the Moyal $\star$ and $\psi
_{A}\left(  x_{0},x_{2n},x_{2n-1}\right)  $ are both independent of
$\mathcal{R}$ and $\mathcal{T}$ \ we see that $\mathcal{R}$ and $\mathcal{T}$
can be removed from the phase space string fields in comparing the left hand
sides of Eqs.(\ref{fourier},\ref{stringFourier}). Therefore we may write the
string field in (\ref{stringFourier}) simply as $A\left(  x_{e},\bar{x}%
,p_{e}\right)  $ and define string field theory using the Moyal star product
given in Eq.(\ref{newstar})
\begin{equation}
\left(  A\star B\right)  \left(  x_{e},\bar{x},p_{e}\right)  .
\end{equation}

The net effect of the intermediate steps involving the split string formalism
with odd modes $l_{2n-1},r_{2n-1}$ is to keep the midpoint $\bar{x}$ fixed
while evaluating string overlaps in Eq.(\ref{wit}). Therefore, in the Moyal
basis $x_{2n},p_{2n}$, the star product of Eq.(\ref{newstar}) must be
evaluated by first writing all string fields $\psi_{A,B}(x_{0},x_{2n}%
,x_{2n-1})$ in terms of $\bar{x}$ instead of $x_{0}$, and then applying the
derivatives with respect to $x_{2n}$. Other than this relic of split strings,
the relation between the original string field $\psi_{A}(x_{0},x_{2n}%
,x_{2n-1})$ and its Fourier transform $A(x_{2n},x_{0},p_{2n})$ given in
Eq.(\ref{fourier}), or the computation of star products, do not involve the
split string formalism.

\section{Split strings with even modes}

It seems puzzling that $\bar{x}$ was distinguished since $x_{0}$ appears to be
more natural in the Moyal basis. Furthermore, $x_{0}$ is gauge invariant under
world sheet reparametrizations, unlike $\bar{x}$. In fact, there is another
split string formalism \cite{abdur} that favors fixing $x_{0}$ rather than
$\bar{x}$ as explained below. First we note the following properties of
trigonometric functions when $0\leq\sigma\leq{\pi}$ for integers $m,n\geq1$
\begin{align}
\cos((2n-1)\sigma)  &  =sign(\frac{\pi}{2}-\sigma)\sum_{m=1}^{\infty}%
[\cos(2m\sigma)-(-1)^{m}]\ \mathcal{T}_{2m,2n-1}\\
\lbrack\cos(2m\sigma)-(-1)^{m}]  &  =sign(\frac{\pi}{2}-\sigma)\sum
_{n=1}^{\infty}\ \cos((2n-1)\sigma)\ \mathcal{R}_{2n-1,2m}. \label{trig}%
\end{align}
Both sides of these equations satisfy Neumann boundary conditions at
$\sigma=0$ and Dirichlet boundary conditions at $\sigma=\frac{\pi}{2}$, and
both are equivalent complete sets of trigonometric functions for the range
$0\leq\sigma\leq\frac{\pi}{2}$ . In the previous section we made the choice of
expanding $l(\sigma),r(\sigma)$ in terms of the odd modes. Now we see that we
could also expand them in terms of the even modes as follows
\begin{equation}
l(\sigma)=\bar{x}+\sqrt{2}\sum_{m=1}^{\infty}\ l_{2m}[\cos(2m\sigma
)-(-1)^{m}]=l_{0}+\sqrt{2}\sum_{m=1}^{\infty}\ l_{2m}\cos(2m\sigma)
\end{equation}
and similarly for $r(\sigma)$. Comparing to the expressions in the previous
section, and using (\ref{trig}) we can find the relation between the odd modes
$(l_{2n-1},r_{2n-1})$ and the even modes $(l_{2n},r_{2n})$
\begin{align}
l_{2n-1}  &  =\sum_{m=1}^{\infty}\ \mathcal{R}_{2n-1,2m}l_{2m},\ \ \ \ l_{2m}%
=\sum_{n=1}^{\infty}\ \mathcal{T}_{2m,2n-1}l_{2n-1}\\
r_{2n-1}  &  =\sum_{m=1}^{\infty}\ \mathcal{R}_{2n-1,2m}r_{2m},\ \ \ \ r_{2m}%
=\sum_{n=1}^{\infty}\ \mathcal{T}_{2m,2n-1}r_{2n-1}%
\end{align}
Furthermore, by using the relation between the odd string modes $(l_{2n-1}%
,\bar{x},r_{2n-1})$ and the full string modes $(x_{0},x_{2n},x_{2n-1})$ in
Eqs.(\ref{l}-\ref{r}) or by direct comparison to $x(\sigma)$, we derive the
relation between the even split string modes and the full string modes.
\begin{equation}
l_{2m}=x_{2m}+\sum_{n=1}^{\infty}\mathcal{T}_{2m,2n-1}x_{2n-1},\qquad
r_{2m}=x_{2m}-\sum_{n=1}^{\infty}\mathcal{T}_{2m,2n-1}x_{2n-1}. \label{even}%
\end{equation}
The inverse relation is
\begin{equation}
x_{2m}=\frac{1}{2}(l_{2m}+r_{2m}),\qquad x_{2m-1}=\sum_{n=1}^{\infty
}\mathcal{R}_{2m-1,2n}\frac{1}{2}(l_{2n}-r_{2n}). \label{evenn}%
\end{equation}
Furthermore, the relations between the zero modes are%
\begin{equation}
l_{0}=x_{0}+\frac{1}{\sqrt{2}}\sum_{n=1}^{\infty}\mathcal{T}_{0,2n-1}%
x_{2n-1},\qquad r_{0}=x_{0}-\frac{1}{\sqrt{2}}\sum_{n=1}^{\infty}%
\mathcal{T}_{0,2n-1}x_{2n-1}.
\end{equation}
Note that the matching condition at the midpoint $l(\pi/2)=r(\pi
/2)=x(\pi/2)=\bar{x}$ is satisfied by the even modes, because $l_{0},r_{0}$
automatically obey the relation
\begin{equation}
l_{0}-r_{0}+\sqrt{2}\sum_{n=1}^{\infty}(l_{2n}-r_{2n})\left(  -1\right)
^{n}=0
\end{equation}
thanks to the property $\mathcal{T}_{0,2n-1}+2\sum_{n=1}^{\infty}%
(-1)^{m}\mathcal{T}_{2m,2n-1}=0$.

This setup allows us to define split string fields that distinguish the center
of mass mode $x_{0}$ rather than the midpoint $\bar{x}$ as follows
\begin{equation}
\hat{A}(l_{2n},x_{0},r_{2n})\equiv\psi_{A}(x_{0},x_{2n},x_{2n-1})=\tilde
{A}(l_{2n-1},\bar{x},r_{2n-1})
\end{equation}
where $(l_{2n},r_{2n})$ are given above in terms of the full string modes. For
this case we define the Witten star product in the form (no integration over
$x_{0}$)
\begin{equation}
\hat{A}\left(  \left\{  l_{2n}\right\}  ,x_{0},\left\{  r_{2n}\right\}
\right)  =\int\,\hat{A}\left(  \left\{  l_{2n}\right\}  ,x_{0},\left\{
z_{2n}\right\}  \right)  \,\hat{B}\left(  \left\{  z_{2n}\right\}
,x_{0},\left\{  r_{2n}\right\}  \right)  \prod_{k}dz_{2k}. \label{witt}%
\end{equation}
Evidently, this product is different than the one in Eq.(\ref{wit}) since the
mode that is held fixed during integration is different (i.e. $x_{0}$ rather
than $\bar{x}$).

We now repeat the arguments that made analogies to section-1 to rewrite this
overlap of half strings in terms of a Moyal product. We find again that the
even modes in phase space $x_{2n},p_{2n}$ are the relevant ones. Furthermore,
the expressions for the string field in phase space $A(x_{2n},x_{0},p_{2n})$
in terms of the original $\psi_{A}(x_{0},x_{2n},x_{2n-1})$ is identical to the
one given in Eq.(\ref{fourier}), and the expression for the star product is
also the same as Eq.(\ref{newstar}). The only difference is that now $x_{0}$
must be held fixed while evaluating the derivatives with respect to $x_{2n}$.
We see that the relic of the split string formalism with even modes is to hold
$x_{0}$ fixed (rather than $\bar{x}$) while performing computations in the
Moyal basis.

More work is required in order to decide which of these procedures is the
correct definition of string field theory. In particular, the symmetries of
the full action, including ghosts, will be relevant in distinguishing them
from each other. Of course, the computation of string amplitudes will also
play a role. We hope to report on further work along these lines in a future publication.

\section{Remarks}

It seems puzzling that only the even modes appear in the Moyal star product.
Although the theory in position space contains both even and odd position
modes $x_{2n},x_{2n-1}$, the mapping of the Witten $\star$ to the Moyal
$\star$ necessarily requires that the Fourier space for the odd positions be
named as the even momenta since $\left(  x_{e}.p_{e}\right)  $ are canonical
under the Moyal star-product. Likewise, the Fourier space for the even
position should be named as the odd momenta. Therefore the double Fourier
transform of the string field $A\left(  x_{e},\bar{x},p_{e}\right)  ,$ with
Fourier kernels of the form (\ref{fourier}) (with $\mathcal{T}$ or
$\mathcal{R}$ as needed) that mix odd-even phase space variables, would be
written purely in terms of odd phase space variables $a\left(  p_{odd},\bar
{x},x_{odd}\right)  .$

In usual phase space quantization all positions and all momenta enter directly
in the Moyal product, however in the present case, which is designed to be
equivalent to Witten's open string field theory, only half of the phase space
variables enter into the definition of the Moyal star product in
Eq.(\ref{newstar}).\footnote{A recent application of deformation quantization
produced the proper approach for discussing two time physics in a field theory
setting. This has conceptual and technical similarities to string field
theory, especially with the new form of string theory based on the Moyal
product in phase space \cite{NCSp}.}

It is straightforward to extend the star product to the ghost sector in the
bosonized ghost formalism. Then the ghost field $\phi(\sigma)$ plays the role
of one extra dimension. If we follow the standard wisdom, our Moyal star
product, including ghosts, would be modified only by inserting a phase at the
midpoint, $exp(\frac{3i}{2}\phi(\frac{\pi}{2}))$, after evaluating the Moyal
product. In view of the discussion in the previous sections one should analyse
this phase insertion more carefully. To construct a string field theory one
would also need to define a BRST operator with the usual properties. The study
of string field theory takes a new form with the new star product. It would be
interesting to see where this leads.

The new Moyal star product in string field theory Eq.(\ref{newstar}) is
Lorentz invariant, in contrast to the Moyal product used in recent studies of
noncommutative field theory (in the presence of a Magnetic field). The string
field formalism includes gauge symmetries that remove ghosts. Since string
theory makes sense, and is ghost free (unitary), it implies that there is at
least one way of making sense of noncommutative field theory when timelike
components of coordinates are included (see also \cite{NCSp} in this respect).

In the presence of a large background $B$-field the midpoint coordinates
$\bar{x}^{\mu}$ become non-commuting as well. In that case, the Moyal star
product in Eq.(\ref{newstar}) is easily modified to accomodate the midpoint
noncommutativity in the usual way. If the $B$-field is small the
noncommutative structure is considerably more complicated.

For the study of $D_{p}$-brane solutions in the vacuum string field theory
approach of \cite{sen}\cite{rsz}\cite{rastsenzwi}\cite{grosstaylor} one seeks
solutions (independent of $x_{0}$ or $\bar{x}$) to the projector equation
\begin{equation}
\left(  A\star A\right)  \left(  x_{e},p_{e}\right)  =A\left(  x_{e}%
,p_{e}\right)  \label{slivmult}%
\end{equation}
in the pure matter sector. Such projectors, involving the Moyal product in
phase space, have been studied for a long time in the literature; they are
known as Wigner functions \cite{wigner} and they have applications in various
branches of physics. It would be simple to generalize the known Wigner
functions to the multi-dimensional string mode space needed in string theory,
and then study their interpretation in string theory. In particular, the
recent solutions obtained in the split string formalism can be easily Fourier
transformed to the phase space formalism. For example, the solution for the
sliver state in our formalism becomes a Gaussian of the form
\begin{equation}
A\left(  x_{e},p_{e}\right)  =(det(2\times1_{e}))^{d}\exp\left(
-x_{e}\mathcal{M}x_{e}-p_{e}\mathcal{M}^{-1}p_{e}\right)  ,\label{sliver}%
\end{equation}
where $d$ is the number of spacetime dimensions, $\mathcal{M}$ is a matrix in
even mode space, $1_{e}$ is the identity matrix in that space, and $det\left(
2\times1_{e}\right)  =\left(  \prod_{n=1}^{\infty}2\right)  $. The phase space
integral over this function gives the rank of the projector, and this is
easily seen to be rank one for any matrix $\mathcal{M}$.
\begin{equation}
TrA=\int\prod_{n=1}^{\infty}\prod_{\mu=0}^{d-1}\frac{dx_{2n}^{\mu}dp_{2n}%
^{\mu}}{2\pi}\ A(x_{e},p_{e})=1.
\end{equation}

In more general computations we anticipate that it would be useful to evaluate
the star product for phase space functions of the form%
\begin{align}
A_{M,\lambda,\mathcal{N}}\left(  x_{e},p_{e}\right)   &  =\mathcal{N\,\,}%
e^{-\eta_{\mu\nu}\left(  x_{e}^{\mu}ax_{e}^{\nu}+x_{e}^{\mu}bp_{e}^{\nu}%
+p_{e}^{\mu}b^{T}x_{e}^{\nu}+p_{e}^{\mu}dx_{e}^{\nu}\right)  -\left(
x_{e}^{\mu}v_{\mu}+p_{e}^{\mu}w_{\mu}\right)  },\quad\label{general}\\
M &  \equiv\left(
\begin{array}
[c]{cc}%
a & b\\
b^{T} & d
\end{array}
\right)  ,\quad\lambda_{\mu}\equiv\left(
\begin{array}
[c]{c}%
v_{\mu}\\
w_{\mu}%
\end{array}
\right)
\end{align}
where in even mode space, $\left(  a,d\right)  $ are symmetric matrices, $b$
is a general square matrix with $b^{T}$ its transpose, $\left(  v_{\mu}%
,w_{\mu}\right)  $ are column matrices, and $\mathcal{N}$ is a normalization
factor. In general these parameters are complex numbers. With the definition
of trace given in Eq.(\ref{trace}), we have%
\begin{equation}
Tr\left(  A_{M,\lambda,\mathcal{N}}\right)  =\frac{\mathcal{N}\,e^{\frac{1}%
{4}\lambda M^{-1}\lambda}}{\left(  \det\left(  2M\right)  \right)  ^{d/2}%
}.\label{Tr}%
\end{equation}
We record the result of our computation of the star product for use in future
applications%
\begin{equation}
\left(  A_{M_{1},\lambda_{1},\mathcal{N}_{1}}\star A_{M_{2},\lambda
_{2},\mathcal{N}_{2}}\right)  \left(  x_{e},p_{e}\right)  =\,A_{M_{12}%
,\lambda_{12},\mathcal{N}_{12}}\left(  x_{e},p_{e}\right)  ,\label{mult}%
\end{equation}
with%
\begin{align}
M_{12} &  =\left(  M_{1}+M_{2}\sigma M_{1}\right)  \left(  1+\sigma
M_{2}\sigma M_{1}\right)  ^{-1}+\left(  M_{2}-M_{1}\sigma M_{2}\right)
\left(  1+\sigma M_{1}\sigma M_{2}\right)  ^{-1}\label{Mrule}\\
\lambda_{12}^{\mu} &  =\left(  1+M_{2}\sigma\right)  \left(  1+M_{1}\sigma
M_{2}\sigma\right)  ^{-1}\lambda_{1}^{\mu}+\left(  1-M_{1}\sigma\right)
\left(  1+M_{2}\sigma M_{1}\sigma\right)  ^{-1}\lambda_{2}^{\mu}%
\label{lammdarule}\\
\mathcal{N}_{12} &  =\mathcal{N}_{1}\mathcal{N}_{2}\left(  \det\left(
1+M_{2}\sigma M_{1}\sigma\right)  \right)  ^{-d/2}e^{\frac{\eta_{\mu\nu}}%
{4}\lambda_{i}^{\mu}K^{ij}\lambda_{j}^{\nu}}\label{Nrule}\\
K^{ij} &  =\left(
\begin{array}
[c]{cc}%
\left(  M_{1}+\sigma M_{2}^{-1}\sigma\right)  ^{-1} & \left(  \sigma
+M_{2}\sigma M_{1}\right)  ^{-1}\\
-\left(  \sigma+M_{1}\sigma M_{2}\right)  ^{-1} & \left(  M_{2}+\sigma
M_{1}^{-1}\sigma\right)  ^{-1}%
\end{array}
\right)  \label{K}%
\end{align}
where $\sigma$ is the purely imaginary matrix that results from the star
commutation rules of $\left(  x_{e},p_{e}\right)  $ in even mode space
$\left[  x_{e}^{\mu},p_{e}^{\nu}\right]  _{\star}=i\eta^{\mu\nu}1_{e}$ and
$\left[  x_{e}^{\mu},x_{e}^{\nu}\right]  _{\star}=0,$ $\left[  p_{e}^{\mu
},p_{e}^{\nu}\right]  _{\star}=0$
\begin{equation}
\sigma=i\left(
\begin{array}
[c]{cc}%
0 & 1_{e}\\
-1_{e} & 0
\end{array}
\right)  .
\end{equation}
Eq.(\ref{mult}) may be used as a generating function for the star products of
more general functions. For example, the star products of more general
functions, such as polynomials multiplied by exponentials of the form
(\ref{general}), can be obtained by taking derivatives of both sides of
(\ref{mult}) with respect to the parameters in $A_{M_{1},\lambda
_{1},\mathcal{N}_{1}},A_{M_{2},\lambda_{2},\mathcal{N}_{2}}.$ We see that
Eqs.(\ref{slivmult},\ref{sliver}) for the projector follow from the more
general multiplication rule (\ref{mult}). We also see that a more general
projector is given by Eq.(\ref{general}) when $M$ satisfies $\sigma
M\sigma=M^{-1}$ and $\mathcal{N}=\left(  \det\left(  2\times1_{e}\right)
\right)  ^{26}\exp\left(  -\lambda^{\mu}M^{-1}\lambda_{\mu}/4\right)  ,$ since
according to (\ref{Mrule}-\ref{K}), one gets $M=M_{1}=M_{2}=M_{3},$ and
$\lambda=\lambda_{1}=\lambda_{2}=\lambda_{12}$ and $\mathcal{N=N}%
_{1}\mathcal{=N}_{2}\mathcal{=N}_{12}.$ The trace of the more general
projector is still $1,$ according to (\ref{Tr}).

It has long been known that Witten's star product defines a noncommutative
geometry for strings, but its relation to other forms of noncommutative
geometry has remained obscure. By making the present bridge to the Moyal star
product one may expect new progress, as well as cross fertilization between
studies in string field theory and noncommutative field theory, and perhaps
even other fields of physics that utilize Wigner functions.

\section*{Acknowledgments}

I would like to thank David Gross, Yutaka Matsuo and Ashoke Sen for useful
discussions, and the ITP M-theory program for their support.

\end{document}